\documentclass[showpacs,onecolumn,aps,prb,floatfix,superscriptaddress,amsmath,amssymb]{revtex4-1}
\usepackage{overpic}
\pdfoutput=1
\usepackage{hyperref}
\usepackage{bookmark}
\usepackage{graphicx}
\usepackage{dcolumn}
\usepackage{bm}
\usepackage{booktabs}
\usepackage{multirow}
\usepackage{color}
\usepackage{soul}
\hypersetup{colorlinks, allcolors={blue}}

\newcommand{\fref} [1]{Fig.~\ref{#1}}

\newcommand{\sref} [1]{Sec.~\ref{#1}}
\newcommand{\Sref} [1]{Section~\ref{#1}}

\newcommand{\aref} [1]{App.~\ref{#1}}

\newcommand{\eref} [1]{Eq.~(\ref{#1})}
\newcommand{\Eref} [1]{Equation~(\ref{#1})}
\newcommand{\eeref}[1]{Eqs.~(\ref{#1})}

\newcommand{\cref} [1]{ref.~\cite{#1}}
\newcommand{\Cref} [1]{Reference~\cite{#1}}

\DeclareMathOperator{\e}{{\displaystyle e}}
\DeclareMathOperator{\de}{{\displaystyle d}}

\DeclareMathOperator{\Li}{{\displaystyle Li}}
\newcommand{\memx}{\mbox{$-\e^{-x}$} } 

\begin{document}

\title{Tutorial  notes for  the evaluation  of thermoelectric  quantum
  bounds in ideal nanostructures}

\author{G. Bevilacqua}
\affiliation{DSFTA, Universit\`{a} di Siena, Via Roma 56, I-53100 Siena, Italy}
\author{A. Cresti}
\affiliation{Univ. Grenoble Alpes, Univ. Savoie Mont Blanc, CNRS, Grenoble INP, IMEP-LAHC, 38000 Grenoble, France}
\author{G. Grosso}
\affiliation{Dipartimento di Fisica ``E. Fermi'', Universit\`{a} di Pisa, Largo Pontecorvo 3, I-56127 Pisa, Italy}
\author{G. Menichetti*}
\affiliation{Dipartimento di Fisica ``E. Fermi'', Universit\`{a} di Pisa, Largo Pontecorvo 3, I-56127 Pisa, Italy}
\affiliation{Center for Nanotechnology Innovation @NEST, Istituto Italiano di Technologia, Piazza San Silvestro 12, 56127 Pisa, Italy.}
\author{G. Pastori Parravicini}
\affiliation{Dipartimento di Fisica ``E. Fermi'', Universit\`{a} di Pisa, Largo Pontecorvo 3, I-56127 Pisa, Italy}
\affiliation{Dipartimento di Fisica ``A. Volta'', Universit\`{a} di Pavia, Via A. Bassi, I-27100 Pisa, Italy}

\date{\today}
\begin{abstract}
  The wave-like  nature of electrons  leads to the existence  of upper
  bounds  on the  thermoelectric  response  of nanostructured  devices
  [R.  S.  Whitney,  Phys.  Rev.   Lett.  {\bf  112},  130601  (2014);
  Phys. Rev. B {\bf 91}, 115425 (2015)].  This fundamental result, not
  present in  classical thermodynamics, was demonstrated  exploiting a
  two-terminal device  modelled by  non-linear scattering  theory.  In
  the present paper, we  consider non-linear quantum transport through
  the  same type  of device  working both  as thermal  machine and  as
  refrigerator.  For  both operations,  starting from charge  and heat
  current expressions,  we provide  analytic quantum bounds  for power
  exchanged, thermal currents and device efficiencies.
  For this  purpose, we adopt  a transmission function  that maximizes
  the  engine efficiency  for  given power  output.   For the  optimal
  boxcar-  or  theta function-transmission  shapes,  we  provide in  a
  tutorial way an explicit deduction  of the quantum bound expressions
  reported in the above cited papers.
\end{abstract}
 
\maketitle
 
\tableofcontents

\section{Introduction}
\label{sec:introduction}

Nanoscale thermoelectric  (TE) engines  for both power  production and
refrigeration have attracted great interest  since the papers of Hicks
and  Dresselhaus~\cite{DRESS93a,DRESS93b,DRESS07}  and  of  Mahan  and
Sofo~\cite{SOFO96}, who evidenced the effect of reduced dimensionality
on the electronic density of states to increase the TE efficiency.  In
linear  regime conditions,  i.e.,  for low  temperature gradients  and
small  applied  voltage  biases,  the  thermoelectric  performance  is
represented  by  the  dimensionless   figure  of  merit  \cite{GOLD99}
$ZT=\sigma  S^2 T/(\kappa_{el}+\kappa_{ph})$,  where  $\sigma$ is  the
electronic conductance, $S$ the  Seebeck coefficient, $T$ the absolute
temperature, and $\kappa_{el}$ ($\kappa_{ph}$) the electronic (phonon)
thermal  conductance.   Attempts  to  reach high  values  of  $ZT$  by
modifying the physical parameters entering in its definition by device
design and  appropriate choice of  materials have shown limits  due to
their     often     competing      behavior     as     function     of
temperature~\cite{DMITRIEV10,DRESS12,NEO15,ZLATIC14,MAHAN16,SNYDER08,NARD16,CULEBRAS14,MASOOD18,URBAN19}.
In  fact,  for nanostructures  the  nonlinear  response regime  is  of
primary interest because  at the nanometer scale  temperature and bias
gradients  may  become  very  large.   Description  of  thermoelectric
phenomena   at   the   nanoscale   as   in   the   case   of   quantum
wells~\cite{MAHAN94},        quantum       dots~\cite{TALBO17,MENI18},
nanowires~\cite{NOZA10},    molecular     junctions~\cite{ZIM17}    or
superlattices~\cite{KARBA16} deserves to  consider fundamental aspects
connected with  quantum effects, thermodynamics and  scale of electron
thermalization~\cite{BookQT,BENENTI17,Jiang2016,Sanchez2016,Sanchez14,LUO18}.
Moreover,   quantum  transport   formalism  provides   an  appropriate
microscopic     description    of     charges    and     heat    flows
~\cite{GOO09,DATTA95,DATTA05}.   In  particular,  for  non-interacting
systems,   calculations   of    thermoelectric   functions   also   in
multiterminal   cases  and   in  the   presence  of   magnetic  fields
~\cite{BRAND13a,BRAND13b}   can    be   done    by   means    of   the
Landauer-B\"{u}ttiker   approach,  which   provides  expressions   for
electron and  heat currents  in terms  of transmission  properties and
contains the microscopic physics of the system.  In the case many-body
effects are important,  the most appropriate approach is  based on the
Keldysh
formalism~\cite{DATTA95,DATTA05,GOO09,CUEVAS10,BALZER13,CRESTI06,FRED05,NIKO12,WANG08}.

\begin{figure}[b]
 \begin{center}
 \includegraphics{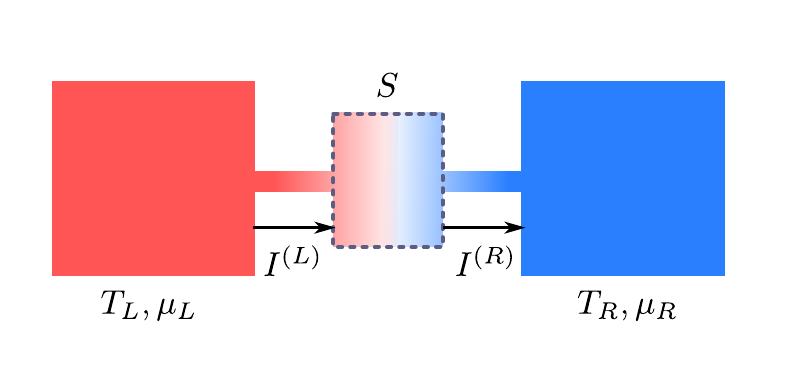}
 \end{center}
 \caption{\label{fig:two-terminal}Schematic   representation  of   the
   scattering  region $S$  connected to  two reservoirs.   We consider
   positive the direction for the currents from the left to right.  }
\end{figure}

In the  present paper, we consider  a two-terminal device made  of two
reservoirs  (left and  right) at  different temperatures  and chemical
potentials,  ($T_L ,\mu_L$)  and ($T_R  ,\mu_R$), and  connected to  a
central     scattering     region     by    perfect     leads,     see
\fref{fig:two-terminal}.
We suppose that the device is  in quantum coherent conditions, with no
electron-phonon  and  electron-electron  interactions.  In  the  above
quantum   coherent   regime,  the   B\"{u}ttiker-Landauer   scattering
theory~\cite{CHRIST96,BENENTI17,SANC13} is used for the description of
heat  currents and  electrical  currents.  This  theory  is valid  for
linear as well for non-linear  regimes.  Transport is described by the
transmission function ${\mathcal T}(E)$ of the scattering region.

A  further  important aspect  connected  with  the quantum  nature  of
electrons has  been highlighted  by Whitney~\cite{WHITNEY14,WHITNEY15}
and addressed  in what follows.   The story goes  back to the  work of
Bekenstein~\cite{BEKE81} on the relation  between information flow and
energy   flow  rates,   and   the  study   of   Lebedev  and   Levitin
\cite{LEBLEV66}(1966)    concerning    the    transmission    of    an
electromagnetic   field  in   one  dimension   and  a   single-channel
communication system.   In a successive  work by means  of information
theory analogy,  Pendry~\cite{PENDRY}(1983) found a  fundamental upper
bound  on the  heat  flow  through a  quantum  system  between a  left
reservoir  at  temperature  $T$  and   a  right  reservoir  at  $T$=0:
$I_Q^{\rm(Pendry)}  \equiv  (k_{\rm   B}T)^2  N  {\pi^2}/{6h}$,  where
$k_{\rm B}$ is  the Boltzmann constant, $N$ is the  number of channels
in  the cross  section through  which current  flows, and  $h$ is  the
Planck constant.

Based on this result, for quantum
thermoelectricity
described by Landauer scattering theory, Whitney extended the Pendry's
result  by  considering the  heat  flow  through a  scattering  system
between  two reservoirs  at  different temperatures  and  at the  same
chemical potential.  He found a quantum  bound on the power output and
then,  by an  optimization process  of the  transmission function,  he
obtained an upper  bound for the efficiency at given  power output for
heat engines and for refrigerators.

In  the present  paper,  we obtain  exactly the  same  results with  a
didactic step-by-step procedure.

  In \sref{sec:model}, we provide basic expressions for thermoelectric
  transport  through  a  two-terminal  mesoscopic  electronic  system.
  These expressions are useful for  the determination of the existence
  of  quantum   bounds  in  currents,  exchanged   power  and  machine
  efficiencies, and for their analytic evaluation.  \Sref{sec:quantum}
  and    \sref{sec:quantum_refrigerators}    present   the    explicit
  expressions for  the above mentioned  quantities.  From them,  it is
  easy to individuate the presence of upper values (quantum bounds) in
  thermoelectric      transport.       \sref{sec:thermal_low}      and
  \sref{sec:refrigerator_low} address  the same above problems  in the
  case  of   very  low  power   exchanged  by  thermal   machines  and
  refrigerators,   respectively.     Finally,   \sref{sec:conclusions}
  concludes.


  \section{Model and basic expressions for thermoelectric transport in
    nanoscale structures} \label{sec:model}
 
  In  this  section,  we  consider transport  through  a  two-terminal
  mesoscopic   electronic   system    coupled   to   two   reservoirs,
  characterized by $N$ transmitting  channels, with total transmission
  function ${\mathcal  T}(E)\leq N$.   Without loss of  generality, we
  assume that  the temperature of  the left reservoir $T_L$  is higher
  than that  of the right reservoir  $T_R$.  We examine in  detail the
  case $\mu_L  < \mu_R$. The  opposite case  $\mu_L > \mu_R$  could be
  envisaged, with appropriate modifications,  from the discussion here
  presented.
 
  The {\it left} or the {\it  right } thermal currents $I_Q^{(L)}$ and
  $I_Q^{(R)}$, and the  output or input power  ${\mathcal P}$, related
  to transport of electrons across  the mesoscopic device are given by
  the   Landauer   expressions   valid  for   linear   and   non-liner
  regimes~\cite{YAMA15,CHRIST96,Sivan1986,Imry1997}
\begin{subequations}
 \label{eq:ini:tot}
\begin{eqnarray}
 I_{Q}^{(L,R)} & = & \frac{1}{h} \int_{-\infty}^{+\infty} dE (E- \mu_{L,R}) 
  \, {\mathcal T}(E) \left[ f_{L}(E) - f_{R}(E) \right] \label{eq:Iq} 
 \\[2mm]
 {\mathcal P} & = & I_{Q}^{(L)} - I_{Q}^{(R)} 
  = \frac{1}{h} \, (\mu_R - \mu_L)
  \int_{-\infty}^{+\infty}
  dE \,
  {\mathcal T} (E) \left[ f_{L}(E) - f_{R}(E) \right] \ , \label{eq:P} 
\end{eqnarray}
\end{subequations}
where  $f_{L,R}(E)=1/[\e^{(E-\mu_{(L,R)})/ (k_{\rm  B}T_{(L,R)})}+ 1]$
are  the Fermi-Dirac  distribution functions  in the  two leads.   The
applied bias potential $\Delta  V=V_L-V_R$ and the chemical potentials
difference    $\Delta   \mu=    \mu_L   -\mu_R$    are   related    by
$(-e)\Delta V  = \Delta \mu  $.  The  device operates as  {\it thermal
  machine},    i.e,   in    the   {\it    power   production    regime
  ${\mathcal P} = {\mathcal P}_{out}  >0$}, when the heat is extracted
from the hot reservoir and released to the cold one, while part of the
thermal energy  can be converted into  usable power. We have  thus the
conditions
\begin{equation} 
 I_Q^{(L)} > I_Q^{(R)} > 0 \ .
\end{equation}
The efficiency of the  device in this mode is defined  as the ratio of
the usable power to the heat extracted from the hot reservoir
\begin{equation}
 \eta^{(tm)} = \frac{{\mathcal P}_{out}}{I_{Q}^{(L)}}
 = \frac{I_{Q}^{(L)}-I_{Q}^{(R)}}{I_{Q}^{(L)}} \le \eta_c^{(tm)} \ .
\end{equation}
As a consequence, the efficiency  of the thermal machine cannot exceed
the  Carnot efficiency  $\eta_c^{(tm)}\equiv (T_L  - T_R  )/T_L$.  The
device operates as  a refrigerator if heat is extracted  from the cold
reservoir and released into the  hot reservoir, with the absorption of
external energy converted  into wasted heat.  In this  case, {\it left
  thermal   current,  right   thermal  current   and  absorbed   power
  (${\mathcal P}={\mathcal P}_{in}$) are all negative quantities}:
\begin{equation} 
 I_Q^{(L)} < I_Q^{(R)} < 0 \ ,
\end{equation} 
and the efficiency (coefficient of performance) of the refrigerator is
given by
\begin{equation}
 \eta^{(refr)} = \frac{ I_{Q}^{(R)} }{{ \mathcal P }_{in}}
 = \frac{ I_{Q}^{(R)} }{ I_{Q}^{(L)} - I_{Q}^{(R)} }
 \le \eta_c^{(refr)} \ .
\end{equation}
As  a  consequence,  the  upper   {\it  thermodynamic  bound}  of  the
refrigeration efficiency  cannot exceed the performance  of the Carnot
refrigerator $\eta_c^{(refr)}\equiv T_R/(T_L-T_R)$.
 
\begin{figure}[tb]
 \begin{center}
 \begin{overpic}[width=0.8\columnwidth]{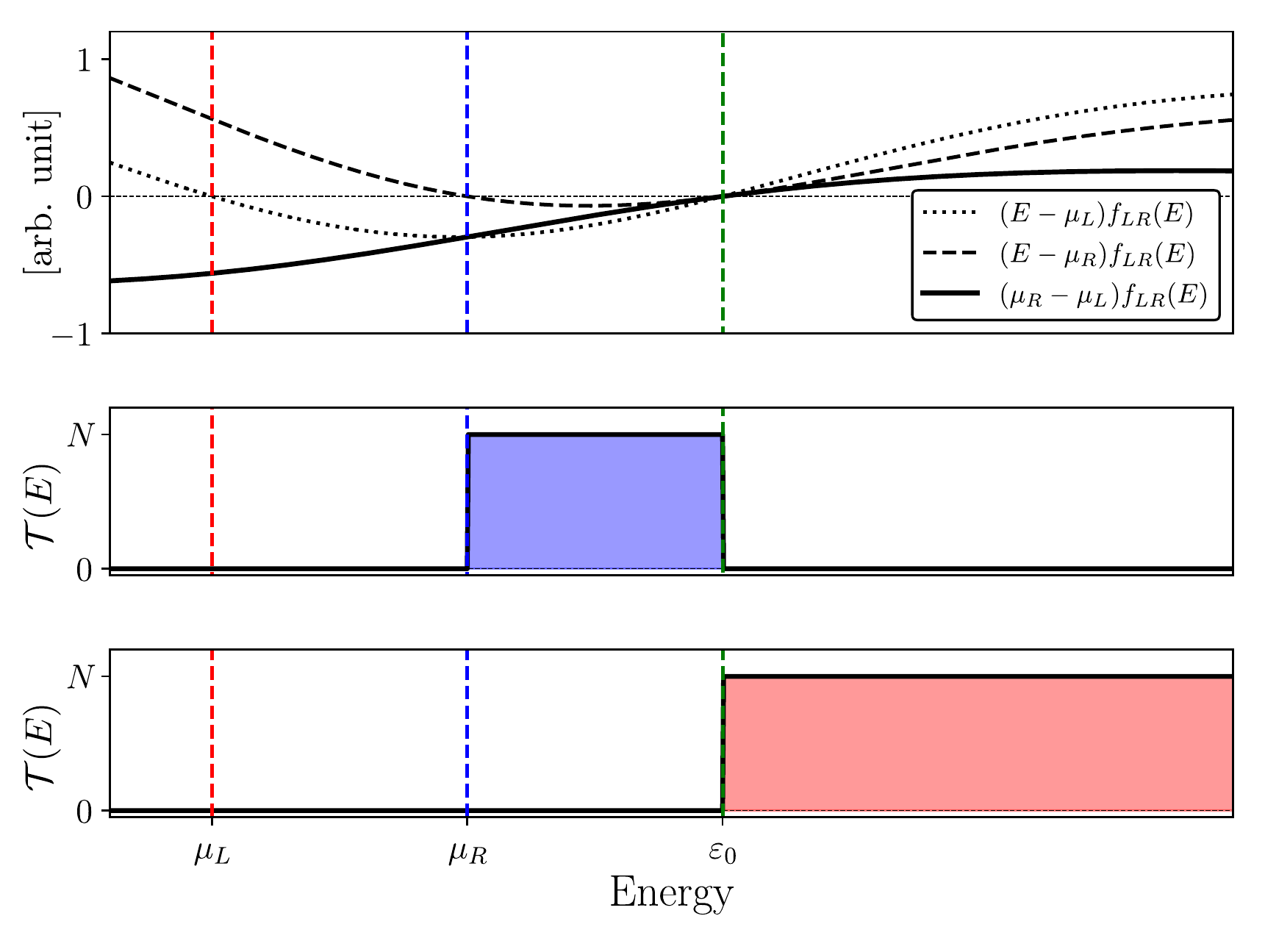}
 \put(0,70){\rm{(a)}}
 \put(0,40){\rm{(b)}}
 \put(0,21){\rm{(c)}}
 \end{overpic}
 \end{center}
 \caption{\label{fig:sign}
   (a)  Representation on  the energy  axis of
   the    functions    $(E    -    \mu_L)f_{LR}(E)$~(pointed    line),
   $(E        -       \mu_R)f_{LR}(E)$~(dashed        line),       and
   $(\mu_R  -   \mu_L)f_{LR}(E)$~(solid  line)  which  enter   in  the
   definition of  the thermal  currents, $I_Q^{(L)}$  and $I_Q^{(R)}$,
   and   power  $\mathcal{P}$   respectively  (see   \eref{eq:Iq}  and
   \eref{eq:P}), in the case $T_L > T_R$ and $\mu_L < \mu_R$.  %
   (b) Optimal  (box-like shape)  transmission function for  the ideal
   refrigerator. It is different from zero, and equal to the number of
   transmission     channels    $N$,     only     in    the     region
   $\mu_R  \leq  E  \leq  \varepsilon_0$ where  both  $I_Q^{(L)}$  and
   $I_Q^{(R)}$ are negative.
   (c) Optimal  (step-like shape) transmission function  for the ideal
   thermal machine. It is different from zero, and equal to the number
   of  transmission channels  $N$, only  when $E  \geq \varepsilon_0$,
   where both $I_Q^{(L)}$ and  $I_Q^{(R)}$ are positive.  Without loss
   of  generality, for  this  figure we  have chosen  $T_L  = 600$  K,
   $T_R = 300$ K, $\mu_L = 0$ eV, and $\mu_R = 0.025$ eV.  }
\end{figure}

From the transport  \eeref{eq:ini:tot}, it is apparent  the basic role
played  by  ${\mathcal T}(E)$  and  by  the  difference of  the  Fermi
functions    of    the    left     and    right    leads    electrodes
$f_{LR}(E)\equiv f_{L}(E) - f_{R}(E)$. We
observe
that
\begin{equation} 
 f_{LR}(E) > 0 \qquad {\rm if} \qquad E > \frac{\mu_R \,T_L - \mu_L \,T_R}{T_L - T_R} \equiv \varepsilon_0 \ ,
\end{equation}
where  the energy  $\varepsilon_0$  separates the  region of  positive
values  of  $f_{LR}(E)$  from  the region  of  negative  values.   The
position  of   $\varepsilon_0$  with  respect  to   the  two  chemical
potentials 
is
\begin{equation} \label{eq:e0-mu}
 \varepsilon_0 - \mu_{L,R} 
 = \frac{T_{L,R}}{T_L - T_R} (\mu_R-\mu_L) > 0 
	\ . 
\end{equation} 
Therefore, $\varepsilon_0$ is at the right of
the  chemical   potentials  of   the  two  reservoirs,   as 
shown in \fref{fig:sign}.

From \eref{eq:e0-mu}, we have 
\begin{equation} 
 \frac{\varepsilon_0 - \mu_L}{k_{\rm B}T_L} = \frac{\varepsilon_0 - \mu_R}{k_{\rm B}T_R} = \frac{\mu_R -\mu_L}{k_{\rm B}(T_L-T_R)} \equiv x_0 \ ,
\end{equation}
where,  as  shown  below,  the  quantity $x_0$  is  important  in  the
definition of quantum bounds for currents and exchanged power.
 
It is worth noticing that at the energy $\varepsilon_0$ the occupation
of  states in  the two  electron reservoirs  is the  same and  the two
reservoirs  can  exchange  electrons   reversibly.   As  evident  from
\eeref{eq:ini:tot},  for   E=  $\varepsilon_0$  the   exchanged  power
${\mathcal  P}$, becomes  $0$  and  the thermal  machine
efficiency reaches the Carnot limit~\cite{HUMP02,HUMP05}:
\begin{equation}
\eta^{(tm) }={\mathcal P}/(I_Q^L )=(\mu_R -\mu_L)/(\varepsilon_0- \mu_L )=1-T_R/T_L =\eta_C^{(tm) } \ .
\end{equation}
This coincides  with the results  of Mahan  and Sofo for  a delta-like
shape transmission  function filtering  at the  energy $\varepsilon_0$
~\cite{SOFO96,BEVI22}  and corresponds  to  reversible transport  with
zero entropy production and  zero output power. Similar considerations
can be  done for  the coefficient of  performance in  the refrigerator
machine.

A  main result of Whitney is the proof that the optimal
efficiency of a  thermal machine at a chosen power  output is obtained
when  the  transmission function  has  a  square shape,  which  allows
transmission of electrons only in a  chosen energy range, the width of
the square being determined by the maximum possible efficiency for the
given power  output~\cite{WHITNEY14,WHITNEY15,Hershfield13},  i.e., a
narrow boxcar for small power output, up to a $\theta$-function shape
for  high power  outputs.   In the  following, we  shall
assume  the above shapes for the ${\mathcal T}(E)$.

\section{Quantum transport through ideal thermal machines}
\label{sec:quantum}

\subsection{Quantum bounds for power generation}
Consider   a   system   in   the   {\it   power   generation   regime}
${\mathcal  P}>0$.  The  expression of  the output  power is  given in
\eref{eq:P}, whose maximal value  is obtained restricting the integral
to  the  positive  region  of $f_{LR}(E)$,  see  \fref{fig:sign},  and
assuming  the value  of  $N$ for  the  total ``optimal''  transmission
function  in  the  whole  domain $[\varepsilon_0,  \infty]$,  i.e.,  a
step-like shape for the transmission  function of the {\it ideal power
  generator.}  For  this device  the  output  power  is given  by  the
expression
\begin{equation}
 \label{eq:Pout}
 {\mathcal P}_{out} = \frac{N}{h} \, (\mu_R - \mu_L) \int_{\varepsilon_0}^{+\infty} dE \, f_{LR}(E) 
 = \frac{N}{h} \, k_{\rm B}^2(T_L-T_R)^2 \, x_0 \ln (1 + e^{-x_0} ) \ ,
\end{equation}
where       we       exploited       the       definite       integral
$\int_{E_0}^{+\infty}   f(E)   \,   dE=    k_{\rm   B}T   \ln   \left[
  1+e^{-(E_0-\mu)/k_{\rm  B}T}  \right]$.   We have  now  to  optimize
\eref{eq:Pout} when the chemical potentials  of the leads are changed,
at fixed temperatures $T_L$ and $T_R $, i.e., to maximize the function
$F(x_0) =  x_0 \ln(1  + e^{-x_0})$.  With  numerical methods,  we find
$dF(x_0)/dx_0=0$ for $\bar x_0= 1.146$ and
$F(\bar x_0)
\cong 0.316$. 
In summary, the quantum bound for power production reads
\begin{equation}
 \label{eq:PoutQB}
 {\mathcal P}_{out}^{(QB)} = C_0 \, \frac{ N}{h} \, k_{\rm B}^2 (T_L-T_R)^2 \ \qquad {\rm with} \qquad C_0 \equiv 0.316 \ ,
\end{equation}
in complete agreement with Eq.(43) and Eq.(44) of \cref{WHITNEY15}.

\subsection{Quantum bounds for left thermal currents}
The  general  expression of  the  left  thermal  current is  given  by
\eref{eq:Iq}.  Its  value  for  a  device  in  the  {\it  ideal  power
  generation  regime}  is obtained  restricting  the  integral to  the
positive region of $f_{LR}(E)$, and  assuming, as before, the value of
$N$  for  the   total  transmission  function  in   the  whole  domain
$[\varepsilon_0, \infty]$. It follows that
\begin{equation} \label{eq:IqL} 
 I_{Q}^{(L)} = \frac{N}{h} \int_{\varepsilon_0}^{+\infty} \!\!\! dE \, (E- \mu_L) f_{LR}(E) \ .
\end{equation}
This integral  is evaluated  analytically  by  using the
elementary properties  of the  poly-logarithm functions of  order one,
${\rm    Li}_1(z)$,   and    of    order    two,   ${\rm    Li}_2(z)$,
(see~\aref{appA:polylogs}).  \Eref{eq:IqL} for  $I_{Q}^{(L)} $ finally
becomes (see~\aref{AppB:general:integrals})
\begin{equation} \label{eq:IqLbis} 
		I_{Q}^{(L)} = \frac{N}{h} \, k_{\rm B}^2 T_L(T_L-T_R) \, x_0 \ln (1+e^{-x_0}) - \frac{N}{h} \, k_{\rm B}^2(T_L^2 - T_R^2) \, {\rm Li}_2(-e^{-x_0}) \ .
\end{equation}
The  thermal  current  depends  on  the  difference  of  the  chemical
potentials, on the difference of the temperatures, on the left current
and on the average temperature.  In the particular case $T_L=T_R$, the
value  of $x_0$  approaches $  \infty$  and no  thermal current  flows
through the device.
 
To establish the quantum bound of the left thermal current with an eye
to \eref{eq:IqLbis}, we have to maximize the function
\begin{equation}
G(x_0) = T_L \, x_0 \ln (1+e^{-x_0}) - (T_L+ T_R) \, {\rm Li}_2(-e^{-x_0}) \ . 
\end{equation}
The   derivative    of   the    above   function,   with    the   help
of~\eqref{sm:eq:polylog:der:exp}, becomes
\begin{equation}
{dG(x_0)}/{dx_0} = - T_L {x_0} / ({e^{x_0} +1} )- T_R \ln(1+e^{-x_0})<0~\ . 
\end{equation}
Since $x_0$  is limited to values  greater or equal to  zero, we argue
that  the   maximum  value  of   the  function  $G(x_0)$   occurs  for
$x_0=0$. By replacing this value into \eref{eq:IqLbis},
we  obtain  that the  left  thermal  current  generated in  the  power
generation regime is limited by the quantum bound
\begin{equation}
 I_{Q}^{(L)(QB)} = \frac{N\pi^2}{12} \, \frac{k_{\rm B}^2}{h} (T_L^2-T_R^2) \ , 
\end{equation}
where we have used ${\rm Li}_2(-1)=-\pi^2/{12}$.

\subsection{Quantum bounds for right thermal currents} 
The maximal value of the right  thermal current of \eref{eq:Iq} in the
power generation regime can be evaluated by following step-by-step the
procedure applied  to the  expression of the  left thermal  current of
\eref{eq:IqL}.   For   the  right  thermal  current,   we  obtain  the
expression
\begin{equation} \label{eq:IqR}
		I_{Q}^{(R)} = \frac{N}{h} \, k_{\rm B}^2 T_R(T_L-T_R) \, x_0 \ln (1+e^{-x_0}) - \frac{N}{h} \, k_{\rm B}^2(T_L^2 - T_R^2) \, {\rm Li}_2(-e^{-x_0}) \ .
\end{equation}
By setting  $\mu_R=\mu_L$, i.e.,  $x_0=0$ in \eref{eq:IqR},  we obtain
that the  maximal value of  the right  thermal current generated  by a
device in the power generation regime presents the quantum bound
\begin{equation}
		I_{Q}^{(R)(QB)} = \frac{N\pi^2}{12} \, \frac{k_{\rm B}^2}{h} (T_L^2-T_R^2) \ .
\end{equation}
This quantum bound is of course the  same as the quantum bound for the
left thermal  current, in fact,  when the two chemical  potentials are
equal, also the two thermal currents must be equal.

\subsection{Quantum bound for efficiency in a power generator}
The expression of the efficiency of the thermal machine reads 
\begin{equation}
 \label{eq:etatm}
 \begin{split}
 \eta^{(tm)} & = \frac{ \mathcal P_{out}}{I_{Q}^{(L)} }
 = \frac{ (T_L-T_R) \, x_0 \ln(1+e^{-x_0}) }
 { T_L \, x_0 \ln(1+e^{-x_0}) - (T_L+T_R) {\rm Li}_2(-e^{-x_0})} \\
 & = \eta_C^{(tm)} \dfrac {1}{1-\dfrac{(T_L+T_R)}{T_L} 
 \dfrac {{\rm Li}_2(-e^{-x_0})} {x_0 \ln(1+e^{-x_0})}} \ .
 \end{split}
\end{equation}

\begin{figure}[h]
 \begin{center}
 \includegraphics[width=0.5\textwidth]{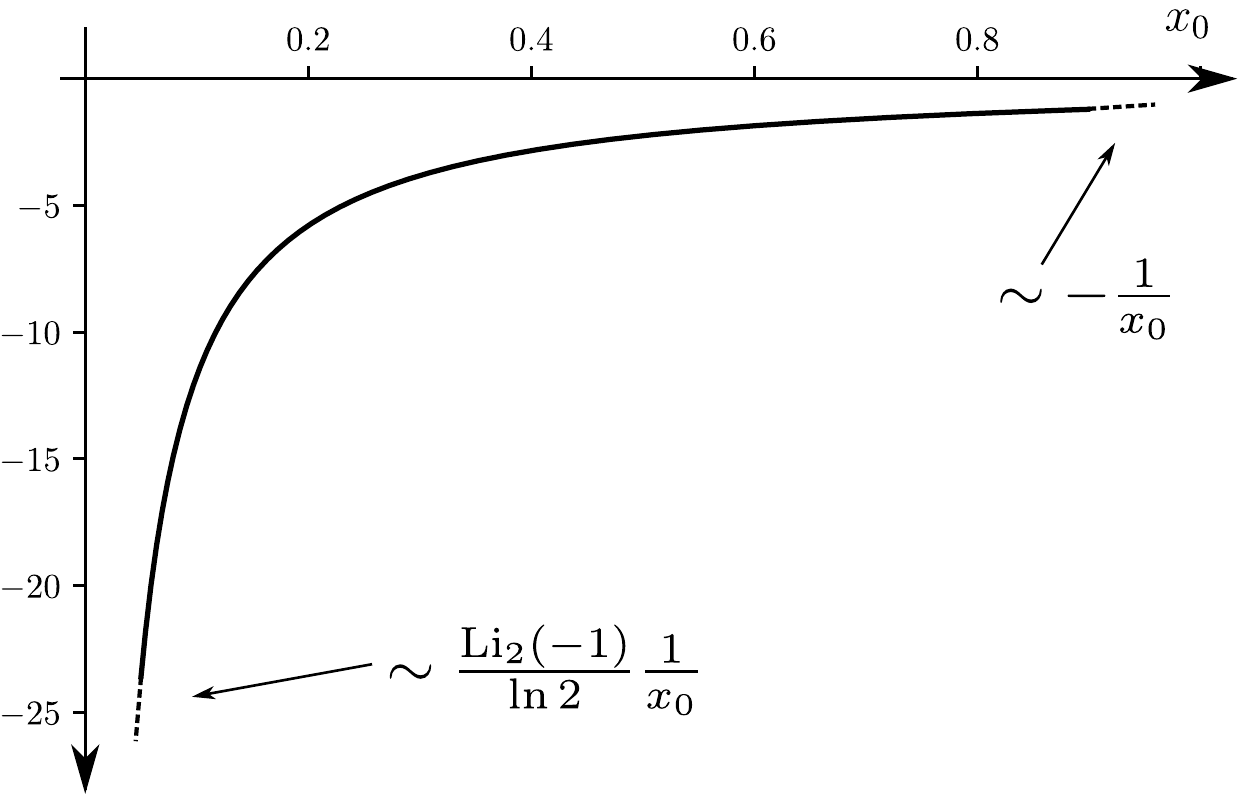}
 \end{center}
 \caption{\label{fig:3} Behavior of the function
 $\dfrac {{\rm Li}_2(-e^{-x_0})} {x_0 \ln(1+e^{-x_0})}$. }
\end{figure}

When  $\mu_L\approx\mu_R$, $\eta^{(tm)}  \approx 0$  and no  efficient
thermal machine is  possible.  On the contrary  the maximal efficiency
of  \eref{eq:etatm} is  obtained  when $x_0  \rightarrow \infty$,  see
Fig.~\ref{fig:3}.   The maximal  efficiency  equals the  thermodynamic
bound  of  the  Carnot  machine,  while the  production  of  power  is
vanishingly small.

\section{Quantum           transport           through           ideal
  refrigerators} \label{sec:quantum_refrigerators}
 
\subsection{Absence of quantum bounds for the absorbed power} 
In  this  section,  we  study transport  through  a  device  perfectly
transparent  in   the  refrigeration   domain  and   perfectly  opaque
elsewhere.  For  a device operating  in the ideal  refrigeration mode,
i.e., with  left thermal current,  right thermal current  and absorbed
power all negative quantities, we restrict the integral defining their
expressions  to  the negative  region  of  $f_{LR}(E)$ in  the  domain
$[\mu_R, \varepsilon_0]$,  and assuming  the ``optimal'' value  of $N$
for the  transmission function  there, see \fref{fig:sign},  i.e., for
the  transmission function  it is  assumed a  box-like shape  of width
$(\varepsilon_0-\mu_R)     $:     ${\mathcal     T}(E)     =N$     for
$\mu_R< E  < \varepsilon_0 $  and ${\mathcal T}(E) =0$  elsewhere.  In
this case, the right thermal current is maximal as requested to have a
maximal   cooling  of   the  right   reservoirs.  For   the  described
nanostructure device in the refrigeration regime (${\mathcal P} < 0$),
the expression of the absorbed (input) power reads
\begin{equation} \label{eq:Pin} 
 {\mathcal P}_{in} = \frac{N}{h} \, (\mu_R - \mu_L) \int_{\mu_R}^{\varepsilon_0} dE \, f_{LR}(E) \ .
\end{equation}
The     above    integral     can     be    performed     analytically
(see~\aref{AppB:general:integrals}) and gives
\begin{equation}
 {\mathcal P}_{in} = \frac{N(\mu_R - \mu_L)}{h} \left[ \! - k_{\rm B}T_L \ln (1\!+\!e^{-x_0})\! +\! k_{\rm B}T_L \ln (1+e^{-x_1}) \!+ \!k_{\rm B}T_R \ln (1\!+\!e^{-x_0})\! - \!k_{\rm B}T_R \ln 2 \right] \ , 
\end{equation}
where $x_1\equiv(\mu_R-\mu_L)/(k_{\rm B}T_L )$.
In particular, we remark 
\begin{equation}
 {\mathcal P}_{in} = - \frac{N(\mu_R - \mu_L)}{h} \, k_{\rm B}T_R \, \ln 2 + \ldots \qquad {\rm for} \qquad \mu_R-\mu_L \, \rightarrow \, + \infty \ .
\end{equation}
For  arbitrary large  difference  of the  chemical  potentials, it  is
evident  that  no  bound  occurs  for the  absorbed  power.   This  is
different from the situation of  thermal machine, where an upper bound
occurs for power generation.
 
\subsection{Left thermal current: absence of quantum bound}
The left thermal current in the ideal refrigerator is
\begin{equation} \label{eq:IqLr}
 I_{Q}^{(L)} = \frac{N}{h} \int_{\mu_R}^{\varepsilon_0} dE \,(E- \mu_L) f_{LR}(E) \ .
\end{equation}
Details  of  the  manipulation  of the  above  equation  are  reported
in~\aref{AppB:general:integrals}. 
The final expression of \eref{eq:IqLr} is 
\begin{equation}
 \label{eq:IQLr:fin}
 \begin{split}
 I_Q^{(L)} = \frac{N}{h} \bigg[ &
 - k_{\rm B}T_R (\mu_R - \mu_L) \ln 2 + k_{\rm B}^2 T_L^2 x_1
 \ln( 1 + \e^{-x_1}) + k_{\rm B}^2 (T_L^2 - T_R^2) \Li_2( - \e^{-x_0} ) \\
 &- k_{\rm B}^2 T_L (T_L - T_R) x_0 \ln( 1+ \e^{-x_0})
 - k_{\rm B}^2 T_L^2\Li_2(-\e^{-x_1}) + k_{\rm B}^2 T_R^2 \Li_2(-1) \bigg] \ .
 \end{split}
\end{equation}
In particular, remembering the definitions of $x_0$ and $x_1$, we find
the leading terms
\begin{equation}
 I_Q^{(L)} = - \frac{N}{h} \, k_{\rm B}T_R \, (\mu_R- \mu_L) \, \ln 2
 + \frac{N}{h} k_{\rm B}^2 T_R^2 \, {\rm Li}_2 (-1) + \ldots 
 \quad {\rm for} \quad \mu_R-\mu_L \, \rightarrow \, + \infty \ .
\end{equation}
For  arbitrary large  difference  of the  chemical  potentials, it  is
evident that  no bound occurs  for the  intensity of the  left thermal
current.  This  is different  from the  situation of  thermal machine,
where an upper bound occurs for the left thermal current.
 
\subsection{Quantum bound for the right thermal current}
By proceeding as above, the right thermal current is
\begin{equation}
 I_{Q}^{(R)} = \frac{N}{h} \int_{\mu_R}^{\varepsilon_0} dE \,(E- \mu_R) f_{LR}(E) \ .
\end{equation}
After some algebra, we obtain the expression
\begin{equation}
 \label{eq:IQRr:fin}
 \begin{split}
 I_Q^{(R)} = \frac{N}{h} \bigg[& k_{\rm B}^2 (T_L^2 - T_R^2) \Li_2( -
 \e^{-x_0} ) - k_{\rm B}^2 T_R (T_L - T_R) x_0 \ln( 1+ \e^{-x_0})\\
 &- k_{\rm B}^2T_L^2\Li_2(-\e^{-x_1}) + k_{\rm B}^2 T_R^2 \Li_2(-1) \bigg] \ .
 \end{split}
\end{equation}
We have that 
\begin{equation}
 I_Q^{(R)} = \frac{N}{h} \, k_{\rm B}^2 T_R^2 \, {\rm Li}_2 (-1) + \ldots
 \quad {\rm for} \quad \mu_R-\mu_L \, \rightarrow \, + \infty \ .
\end{equation}
In conclusion, a  quantum bound exists for the  negative right thermal
current
\begin{equation} \label{eq:IQLbound}
 \left|I_{Q}^{(R)}\right| < \frac{N\pi^2}{12} \, \frac{k_{\rm B}^2}{h} \, T_R^2 \equiv I_Q^{R(QB)} 
 \ . 
\end{equation}
The above  result is  in agreement  with that  reported in  Eq.(58) of
\cref{WHITNEY15}, considering that we define 
\begin{equation}
 I_Q^{R(QB)}
 = \frac{1}{2} \, I_Q^{\rm (Pendry)} \ .
\end{equation}

\section{Quantum transport in a thermal machine at low-power output.} \label{sec:thermal_low}
We  consider  quantum transport  through  a  device in  the  low-power
generation regime, i.e., in the  case ${\mathcal P} \approx 0^+$.  The
ideal low-power generation regime is obtained restricting the integral
in  Eq.~\eqref{eq:P}  to  the  positive  region  of  $f_{LR}(E)$,  and
assuming the value  of $N$ for the transmission function  in the small
domain $[\varepsilon_0, \varepsilon_0 + \Delta]$ (i.e., a {\it box-car
  shape} of  width $\Delta$, where  we expect maximum  efficiency, see
\fref{fig:two-terminal}).  Namely, we assume ${\mathcal T}(E) = N$ for
$\varepsilon_0    <    E    <   \varepsilon_0    +    \Delta$,    with
$\Delta \rightarrow 0^+$, and ${\mathcal T}(E) =0$ otherwise.
The low-power output becomes
\begin{equation}
 {\mathcal P}_{out} = \frac{N}{h} \, (\mu_R - \mu_L) \int_{\varepsilon_0}^{\varepsilon_0 + \Delta} dE \, f_{LR}(E) \ .
\end{equation} 
Proceeding as in~\aref{AppC:infinitesimal:integral}, 
one can show that the final result is 
\begin{equation}
 \label{eq:Poutl} 
 {\mathcal P}_{out} = \frac{N}{h} (\mu_R - \mu_L) \left[
 \frac{\Delta^2 (T_L-T_R)}{2k_{\rm B}T_LT_R} \psi'(x_0) +\frac{\Delta^3(T_L^2 -T_R^2)}{6k_{\rm B}^2T_L^2T_R^2} \psi''(x_0) + O(\Delta^4) \right] \ , 
\end{equation}
where $\psi(x) = -1/(e^x +1)$.
 
\subsection{Left thermal current in low-power ideal generators}
Following~\aref{AppC:infinitesimal:integral}, the left thermal current is
\begin{equation}
 \label{eq:IqLpg}
 \begin{split}
 I_{Q}^{(L)} &= \frac{N}{h} \int_{\varepsilon_0}^{\varepsilon_0+\Delta} dE \, (E- \mu_L) f_{LR}(E)\\ 
 & = \frac{N}{h} (\mu_R - \mu_L) \frac{T_L}{T_L - T_R} \left[ \frac{\Delta^2 (T_L-T_R)}{2k_{\rm B}T_LT_R} \, +\psi'(x_0) + \frac{\Delta^3 (T_L^2-T_R^2)}{6k_{\rm B}^2 T_L^2T_R^2} \, \psi''(x_0) \, \right] \\
 &\phantom{=} + \frac{N}{h} \frac{\Delta^3 (T_L-T_R)}{3k_{\rm B}T_LT_R} \, \psi'(x_0) +O(\Delta^4)
 \ .
\end{split}
\end{equation}
By using \eref{eq:Poutl}, we obtain the more effective form
\begin{equation}
 \label{eq:IQLeff} 
 I_Q^{(L)} = \frac{T_L}{T_L - T_R} \, {\mathcal P}_{out} 
 + \frac{N}{h} \frac{\Delta^3 (T_L-T_R)}{3k_{\rm B}T_LT_R} \, \, \psi'(x_0)
 + O(\Delta^4) \ ,
\end{equation}
which exactly coincides with Eq.(47) of \cref{WHITNEY15}.

\subsection{Efficiency of low-power thermal machines}
 
We can divide both members of \eref{eq:IQLeff} by $I_Q^{(L)}$ and obtain 
\begin{equation}
\label{eq:prova}
 \eta^{(tm)} = \frac{ {\mathcal P}_{out} }{I_Q^{(L)} }
 = \frac{T_L-T_R}{T_L} \left[ 1 
 - \frac{N}{h} \frac{\Delta^3 (T_L-T_R)}{3k_{\rm B}T_LT_R} \, 
 \psi'(x_0) \frac{1}{I_Q^{(L)} } \right] .
\end{equation}
Linear  corrections   in  the  $\Delta$  parameter   can  be  obtained
expressing   $I_Q^{(L)}$  to   quadratic   terms   in  $\Delta$   from
\eref{eq:IqLpg}. Eventually, we find
\begin{equation}
 \label{eq:etatmbis} 
 \eta^{(tm)} = \eta_c^{(tm)} \left[ 1 - \frac{2}{3} \, \frac{\Delta}{k_{\rm B}T_L x_0} + O(\Delta^2) \right] 
\end{equation}
thus recovering Eq.(49) of \cref{WHITNEY15}.

\subsection{Efficiency of low-power thermal machines at fixed power output}
The  expression for  the  power production  of  the low-power  thermal
machine is given  by \eref{eq:Poutl}. At the  lowest (quadratic) order
in $\Delta$, the usable power reads
\begin{equation} 
 {\mathcal P}_{out} = \frac{N}{h} \, x_0 \, \frac{\Delta^2 (T_L-T_R)^2}{2T_LT_R} \, \psi'(x_0) \ .
\end{equation} 
It is convenient to normalize the  output power with the quantum bound
obtained in \eref{eq:PoutQB}, i.e.,
\begin{equation} 
 \frac {{\mathcal P}_{out}}{{\mathcal P}_{out}^{(QB)}} 
 = \frac{1}{C_0} \, \frac{\Delta^2 }{2k_{\rm B}^2T_LT_R} \, x_0 \psi'(x_0) \ .
\end{equation}
By combining \eref{eq:etatmbis} with the above equation, we obtain
\begin{equation} \label{eq:etaTM} 
 \eta^{(tm)} = \eta_c^{(tm)} \left[ 1 - \frac{2}{3} \, 
 \sqrt{ 2C_0 } \, \sqrt{ \frac{T_R}{T_L} \,
 \frac {{\mathcal P}_{out} } {{\mathcal P}_{out}^{(QB)} } } 
 \, \frac{1}{ \sqrt{ x_0^3 \psi'(x_0)} } \right] .
\end{equation}
The  last step  to  be  performed is  the  maximization  of the  above
expression.

\subsection{Optimization of the efficiency of low-power thermal machine at fixed power output}
Optimization  of the  efficiency  at given  temperatures requires  the
maximization of  the function  $H(x_0) =  x_0^3 \,  \psi'(x_0)$, which
appears in the denominator of \eref{eq:etaTM}.  With standard methods,
we find
\begin{equation}
 \label{eq:x0bar} 
 \bar x_0 =3.24 \ \ \ {\rm and} \ \ \ H(\bar x_0) = 1.234 \ .
\end{equation}
Since, according to \eref{eq:PoutQB}, $C_0= 0.316$, we have
  $(2/3)\sqrt{ {2C_0}/{H(\bar  x_0)}}=0.477$  and finally
\begin{equation} 
 \eta^{(tm)} = \eta_c^{(tm)} \left[ 1 - 0.477 \, \sqrt{ \frac{T_R}{T_L} } 
 \, \sqrt{ \frac{ {\mathcal P}_{out} }
 { {\mathcal P}_{out}^{(QB)} } } \ \right] \ . 
\end{equation}
This relation coincides with Eq.(51) of \cref{WHITNEY15}.

\section{Quantum   transport   in   a  refrigerator   at   low-cooling
  regime} \label{sec:refrigerator_low}
The  ideal  low-power refrigerator  (${\mathcal  P}  \approx 0^-)$  is
obtained  restricting   the  integral   to  the  negative   region  of
$f_{LR}(E)$,  and  assuming the  value  of  $N$ for  the  transmission
function            in           the            small           domain
$[\varepsilon_0  - \Delta,  \varepsilon_0]$, where  we expect  maximum
efficiency, see \fref{fig:sign}, i.e., we assume ${\mathcal T}(E) = N$
for   $\varepsilon_0    -   \Delta   <   E    <   \varepsilon_0$   and
$ {\mathcal T}(E) = 0$ otherwise, with $ \Delta \rightarrow 0^+$.  For
the low-cooling refrigerator, the absorbed power is then
\begin{equation}
 {\mathcal P}_{in} = \frac{N}{h} \, (\mu_R - \mu_L) \int_{\varepsilon_0 - \Delta}^{\varepsilon_0} dE \, f_{LR}(E) \ .
\end{equation}
With   an   eye   to~\aref{AppC:infinitesimal:integral},   the   power
absorption becomes
\begin{equation}
 \label{eq:Pinab}
 {\mathcal P}_{in} = \frac{N}{h}
 (\mu_R - \mu_L) \left[ -\frac{\Delta^2 (T_L-T_R)}{2k_{\rm B}T_LT_R}
 \psi'(x_0) + \frac{\Delta^3(T_L^2 -T_R^2)}{6k_{\rm B}^2T_L^2T_R^2}
 \psi''(x_0) \right] + O(\Delta^4) \ .
\end{equation}
%
 
\subsection{Right thermal current for an ideal refrigerator in the low-cooling regime}
In the present conditions, the right thermal current is
\begin{eqnarray}
 I_{Q}^{(R)} & = & \frac{N}{h} \int_{\varepsilon_0 - \Delta}^{\varepsilon_0} dE \, (E- \mu_R) f_{LR}(E) \nonumber \\[3mm] 
 & = & \frac{N}{h} (\mu_R - \mu_L) \frac{T_R}{T_L - T_R}
 \left[ - \frac{\Delta^2 (T_L-T_R)}{2k_{\rm B}T_LT_R} \, \psi'(x_0)
 + \frac{\Delta^3 (T_L^2-T_R^2)}{6k_{\rm B}^2 T_L^2T_R^2}
 \, \psi''(x_0) \, \right] \nonumber \\[3mm]
 &&
 + \frac{N}{h} \frac{\Delta^3 (T_L-T_R)}{3k_{\rm B}T_LT_R} \, \,
 \psi'(x_0) + O(\Delta^4) \ .
 \label{eq:IqRab} 
\end{eqnarray}
By exploiting \eref{eq:Pinab} for ${\mathcal P}_{in}$, \eref{eq:IqRab}
can be cast in the more effective form
\begin{equation}
 I_Q^{(R)} = \frac{T_R}{T_L - T_R} \, {\mathcal P}_{in} + \frac{N}{h} \frac{\Delta^3 (T_L-T_R)}{3k_{\rm B}T_LT_R} \, \, \psi'(x_0) \ .
\end{equation}
This equation coincides with  Eq.(59) of \cref{WHITNEY15}, where $T_R$
and $T_L$ are exchanged.

\subsection{Efficiency of an ideal refrigerator in the low-cooling regime}
The general expression for the  efficiency of the refrigerator machine
reads
\begin{equation} \label{eq:etarfr}
 \eta^{(refr)} = \frac{\ \ I_Q^{(R)} } { {\mathcal P}_{in} }
 = \frac{T_R}{T_L-T_R} \left[ 1 + \frac{N}{h} \frac{\Delta^3 (T_L-T_R)^2}{3k_{\rm B}T_LT_R^2} \, \psi'(x_0) \frac{1} { {\mathcal P}_{in} } \right] .
\end{equation}
Correction linear in the $\Delta$ parameter can be obtained expressing
${\mathcal   P}_{in}$   to   quadratic  terms   in   $\Delta$.    From
\eref{eq:Pinab} it holds
\begin{equation} \label{eq:Prfr}
 {\mathcal P}_{in} = -\frac{N}{h} (\mu_R - \mu_L) \frac{\Delta^2 (T_L-T_R)}{2k_{\rm B}T_LT_R} \, \psi'(x_0) + { O}(\Delta^3) 
\end{equation}
Inserting \eref{eq:Prfr} into \eref{eq:etarfr} gives 
\begin{equation}
 \eta^{(refr)} = 
 \eta_c^{(refr)} \left[ 1 - \frac{2}{3} \, \frac{\Delta}{k_{\rm B}T_R x_0} +
 O(\Delta^2)
 \right],
\end{equation}
which corresponds to Eq.(61) of \cref{WHITNEY15}. 

\subsection{Efficiency of the low-cooling refrigerator at fixed power input}
The  expression  for  the  right  thermal  current  of  the  low-power
refrigeration  machine is  given  by \eref{eq:IqRab}.   At the  lowest
(quadratic) order in $\Delta$, the right thermal current reads
\begin{equation} 
 I_Q^{(R)} = -\frac{N}{h} \, x_0 \, \frac{\Delta^2 (T_L-T_R)}{2T_L} \, \psi'(x_0) \ .
\end{equation}
%
As done in Eq.(20) of \cref{WHITNEY15}, we normalize the right thermal
current as
\begin{equation} 
 \frac{I_Q^{(R)}}{I_Q^{\rm (Pendry)}} = \frac{1}{2C_1} \, \frac{\Delta^2 (T_L-T_R)}{k_{\rm B}^2T_LT_R^2} \, x_0 \psi'(x_0)
 \ \ \ \ \ {\rm with} \ \ \ \ \ C_1=\frac{\pi^2}{6}
 \ .
\end{equation}
The expression of $\Delta$ from the  above equation can be inserted in
Eq.~\eqref{eq:etarfr} and one obtains
\begin{equation} \label{eq:etarfr2} 
 \eta^{(refr)} = \eta_c^{(refr)} \left[ 1 - \frac{2}{3} \, \sqrt{ 2C_1 } \, 
 \sqrt{\frac{T_L}{T_L-T_R} \, \frac{I_Q^{(R)}}{I_Q^{\rm (Pendry)}}} 
 \, \frac{1}{\sqrt{x_0^3 \psi'(x_0)}} \right] .
\end{equation}
The  last step  to  be  performed is  the  maximization  of the  above
expression.

\subsection{Optimization   of   the   efficiency  of   the   low-power
  refrigerator at fixed power absorption}

Optimization  of  the efficiency  at  fixed  temperatures requires  in
essence       the       maximization       of       the       function
$H(x_0)  =  x_0^3  \,  \psi'(x_0)$, with  $H(x_0)>0$.   This  function
appears in  the denominator  of \eref{eq:etarfr2}.   This can  be done
repeating  step by  step the  same process  used for  the case  of the
low-power    thermal    machine    at   fixed    power    output    in
Section~\ref{sec:thermal_low}.  In particular, by exploiting the value
of     the     coefficient     $C_1     =     1.645$     we     obtain
$(2/3) \sqrt{{2C_1}/{H(x_0)}} = 1.088$.  Finally, the maximal value of
Eq.~\eqref{eq:etarfr} is
\begin{equation} 
 \eta^{(refr)} = \eta_c^{(refr)} \left[ 1 - 1.088 \, \sqrt{\frac{T_L}{T_L-T_R} \, \frac {I_Q^{(R)}}{I_Q^{\rm (Pendry)}}} \right] \ ,
\end{equation}
which recovers exactly Eq.(63) of \cref{WHITNEY15}.

\begin{table}[t]
\begin{center}
\begin{tabular}{|c|c|c|}
\hline
 \rule[-7mm]{0mm}{1.6cm}
 & Ideal thermal machine & Ideal refrigerator \\
 \hline
 \hline
 \rule[-9mm]{0mm}{2.1cm}
$\mathcal{T}(E)$ & $\mathcal{T}(E):= \begin{cases}
 N,\quad E\geq \varepsilon_0\\
 0,\quad E <\varepsilon_0 
\end{cases}$& $\mathcal{T}(E):= \begin{cases}
 N,\quad \mu_R \leq E \leq \varepsilon_0\\
 0,\quad {\rm Otherwise}
\end{cases}$ \\
 \hline
 \rule[-7mm]{0mm}{1.6cm}
 $\mathcal{P}^{(QB)} $ & $\mathcal{P}^{(QB)}_{out}=C_0\dfrac{N}{h}k_{\rm B}^2(T_L-T_R)^2$ & Absence of bound for $\mathcal{P}_{in} $\\
 \hline
 \rule[-7mm]{0mm}{1.6cm}
 $I^{(L)(QB)}_Q$ & $ \dfrac{N\pi^2}{12}\dfrac{k_{\rm B}^2}{h}(T^2_L-T^2_R)$ & Absence of bound for $I^{(L)}_Q$ \\
 \hline
 \rule[-7mm]{0mm}{1.6cm}
 $I^{(R)(QB)}_Q$ & $ \dfrac{N\pi^2}{12}\dfrac{k_{\rm B}^2}{h}(T^2_L-T^2_R)$ & $\dfrac{N\pi^2}{12}\dfrac{k^2_{\rm B}}{h}T^2_R$\\
 \hline
 \rule[-7mm]{0mm}{1.6cm}
 $\eta$ & $\eta^{(tm)}\rightarrow\quad \eta_{\rm C}^{(tm)}$ & Absence of bound for $\eta^{(refr)}$ \\
 \hline

 \hline
 
 \end{tabular}
 \caption{
 Transmission functions, quantum bounds and efficiency for the ideal thermal machine and for the ideal refrigerator.
}
 \label{tab:QB1}
 \end{center}

\end{table}

\begin{table}[t]
\begin{center}
\begin{tabular}{|c|c|}
\hline
 \rule[-7mm]{0mm}{1.6cm}
 Low-power output ideal thermal machine & Refrigerator in the low-cooling regime \\
 \hline
 \hline
 \rule[-9mm]{0mm}{2.1cm}
 $\mathcal{T}(E):=\begin{cases}
 N,\quad \varepsilon_0 \leq E\leq \varepsilon_0 + \Delta, \quad \Delta\rightarrow 0^+\\
 0,\quad {\rm Otherwise}
\end{cases}$ & $\mathcal{T}(E):= \begin{cases}
 N,\quad \varepsilon_0-\Delta \leq E \leq \varepsilon_0, \quad \Delta\rightarrow 0^+\\
 0,\quad {\rm Otherwise}
\end{cases} $ \\ 
 \hline
 \rule[-7mm]{0mm}{1.6cm}
 $\eta^{(tm)}\sim\eta_{\rm C}^{(tm)}\left[\-0.477\sqrt{\dfrac{T_R}{T_L}}\sqrt{\dfrac{\mathcal{P}_{out}}{\mathcal{P}^{(QB)}_{out}}} \right]$ & $\eta^{(refr)}\sim\eta_{\rm C}^{(refr)}\left[1-1.088\sqrt{\dfrac{T_L}{T_L-T_R}\dfrac{I^{(R)}_Q}{I^{(R)(QB)}_Q}} \right]$ \\
 \hline
 \hline
\end{tabular}
\caption{
 Transmission functions and maximum efficiencies for the ideal thermal
 machine  and  the  refrigerator  in  the  low-power  and  low-cooling
 regimes, respectively. 
}
 \label{tab:QB2}
 \end{center}
\end{table}

\section{Conclusions}
\label{sec:conclusions}

We have presented  the analytic details for the  evaluation of quantum
bounds in the response functions of thermoelectric machines.

We have considered a two-terminal macroscopic device operating between
two reservoirs at different temperatures and chemical potentials.  The
electronic  transport  in  the  device is  modelled  by  the  Landauer
scattering   theory   and   the   electronic   transmission   function
${\mathcal  T}(E)$ is  assumed to  be of  step-like shape  or box-like
shape,  in  the  case  of power  generation  machine  or  refrigerator
machine, respectively,  so to guarantee  the condition of  upper bound
value of the thermoelectric  parameters.  The expressions obtained are
summarized in tables~\ref{tab:QB1} and~\ref{tab:QB2}.

We hope that  the present analysis of currents and  quantum bounds can
be useful for highlighting the  formal aspects of the results obtained
in the literature.

\section*{Acknowledgments}
G. M. acknowledges support from the  University of Pisa under the “PRA
- Progetti  di Ricerca  di Ateneo”  (Institutional Research  Grants) -
Project No. PRA 2020-2021 92.
\appendix

\section{ Polylogarithms  basic properties}  \label{appA:polylogs} The
poly-logarithm  function of  order unit  is  defined in  terms of  the
standard logarithm as~\cite{MR618278}
\begin{equation}
 \label{sm:eq:polylog:1}
 \Li_1 (z) \equiv -\ln(1-z) = \sum_{1}^{\infty} \frac{z^n}{n} \ ,
\end{equation}
$|z|<1$.
In general, one defines the polylog of order $k$ as
\begin{equation}
 \label{sm:eq:polylog:k}
 \Li_k(z) \equiv \int_0^z \frac{\Li_{k-1}(t)}{t} \de t = 
 \sum_1^{\infty}\frac{z^n}{n^k} \qquad k=2,3,\ldots
\end{equation}
where  again the  series expansion  is  valid in  the $|z|<1$  region.
These functions inherit  from the logarithm the branch  point at $z=1$
and the cut along the positive real axis $[1, +\infty)$.
Considering $z= - \e^{-x}$ one obtains 
\begin{equation}
 \label{sm:eq:polylog:der:exp}
 \frac{d}{d x} \Li_{m}(-\e^{-x}) = - \Li_{m-1}(\memx) \qquad m=1,2,\ldots
\end{equation}
Notice in particular that
\begin{equation}
 \label{sm:eq:polylog:Li0}
 \Li_0(\memx) = - \frac{d}{dx} \Li_1(\memx) = - \frac{1}{\e^x + 1} \ .
\end{equation}
The above  functions are useful  for calculating integrals  related to
the Fermi functions.  In fact
\begin{equation}
 \label{sm:eq:polylog:int:0}
 \int \frac{1}{\e^x + 1} \de x = - \int \Li_0(\memx) \de x =
 \Li_1(\memx) = -\ln( 1 + \e^{-x})
\end{equation}
and integrating by parts
\begin{equation}
 \label{sm:eq:polylog:int:1}
 \begin{split}
 \int \frac{x}{\e^x + 1} \de x 
 & = \int x \frac{d}{d x} \Li_1(\memx) \, \de x 
 = x \Li_1(\memx) - \int \Li_1(\memx) \de x\\
 &= -x \ln(1 + \e^{-x}) + \Li_2(\memx) \ .
\end{split}
\end{equation}

Asymptotic behaviors~\cite{MR618278} for $  x \rightarrow +\infty$ can
be obtained easily as
\begin{equation}
 \label{sm:eq:asym:destra}
 \Li_1 (\memx) = \memx + O( \e^{-2x} ) \qquad \mathrm{and} \qquad
 \Li_2 (\memx) = \memx + O( \e^{-2x} ) \ ,
\end{equation}
moreover, for $x\rightarrow -\infty$ 
\begin{equation}
 \label{sm:eq:asym:sinistra}
 \Li_1 (\memx) = x - \e^x + O( \e^{2x} ) \qquad \mathrm{and} \qquad
 \Li_2 (\memx) = -\frac{\pi^2}{6} - \frac{x^2}{2} + \e^x + O( \e^{2x} ) \ .
\end{equation}

\section{Currents             and            power             related
  integrals}
\label{AppB:general:integrals}
Let us introduce two primitive functions involving the Fermi function
\begin{equation}
 \label{sm:eq:primitive:ff}
 g(E) \equiv \int \frac{1}{\e^{\beta (E - \mu)}+1 }\de E
 = \frac{1}{\beta}\Li_1( - \e^{-\beta (E - \mu)})
\end{equation}
and
\begin{equation}
 \label{sm:eq:primitive:Eff}
 h(E) \equiv \int \frac{E}{\e^{\beta (E - \mu)}+1 }\de E
 = E g(E) + \frac{1}{\beta^2}\Li_2( - \e^{-\beta (E - \mu)}) \ ,
\end{equation}
which      follow       from      \eqref{sm:eq:polylog:int:0}      and
\eqref{sm:eq:polylog:int:1}, respectively, and  $\beta=1/k_{\rm B} T$.
Notice  that  both $g(E)$  and  $h(E)$  are exponentially  small  when
$E\rightarrow +\infty$.

With      the     help      of     \eqref{sm:eq:primitive:ff}      and
\eqref{sm:eq:primitive:Eff}  we  can  quickly  express  the  integrals
reported in the main text.  In fact
\begin{subequations}
 \label{sm:eq:int:gh}
\begin{align}
 \label{sm:eq:int:g}
 \int f_{LR}(E) \de E & = g_L(E) - g_R(E)\\
 \label{sm:eq:int:h}
 \int (E-\mu) f_{LR}(E) \de E
  &=
  h_L - h_R - \mu(g_L - g_R) \nonumber\\
  &=(E-\mu)(g_L(E) - g_R(E)) \nonumber \\
  &\phantom{=} + \beta_L^{-2}
  \Li_2(-\e^{-\beta_L (E - \mu_L )})
  -\beta_R^{-2} \Li_2(-\e^{-\beta_R (E - \mu_R )}) \ ,
\end{align}
\end{subequations}
where $g_{L,R}(E)$ mean that $g(E)$ is calculated in correspondence of
$T_{L,R}$   and   $\mu_{L,R}$.   From   application   of   expressions
\eqref{sm:eq:int:gh} we recover the results reported in the main text.
For example, for the left current  in the thermal machine mode we need
to evaluate
\begin{equation}
 \label{sm:eq:IQL:tm}
 \begin{split}
 \int_{\epsilon_0}^{+\infty} (E-\mu_L) f_{LR}(E) \de E & =
 -(\epsilon_0-\mu_L)(g_L(\epsilon_0) - g_R(\epsilon_0)) +\\
 &\phantom{=} - \beta_L^{-2} \Li_2(-\e^{-\beta_L (\epsilon_0 - \mu_L )})
 +\beta_R^{-2} \Li_2(-\e^{-\beta_R (\epsilon_0 - \mu_R )}) \\
 &= k_{\rm B}^2T_L(T_L- T_R) \, x_0\, \ln( 1 + \e^{-x_0}) -
 k_{\rm B}^2(T_L^2 - T_R^2)\Li_2(-\e^{-x_0} ) \ ,
\end{split}
\end{equation}
where      we      exploited      the      asymptotic      expressions
\eqref{sm:eq:asym:destra} and the definition of $\epsilon_0$ and $x_0$
given in the main text.

In the  refrigerator mode the  expressions are more  involved, because
both integration limits are finite. For the left current we need
\begin{equation}
 \label{sm:eq:IQL:refr}
 \begin{split}
 \int_{\mu_r}^{\epsilon_0} (E-\mu_L) f_{LR}(E) \de E & =
 (\epsilon_0-\mu_L)(g_L(\epsilon_0) - g_R(\epsilon_0))
 -(\mu_R-\mu_L)(g_L(\mu_R) - g_R(\mu_R)) \\
 &\phantom{=} + \beta_L^{-2} [ \Li_2(-\e^{-x_0}) - \Li_2(-\e^{-x_1})]
 -\beta_R^{-2} [\Li_2(-\e^{-x_0}) - \Li_2(-1)] \ .
 \end{split}
\end{equation}
After some algebra, one obtains
\begin{equation}
 \label{sm:eq:IQL:refr:fin}
 \begin{split}
 \int_{\mu_r}^{\epsilon_0} (E-\mu_L) f_{LR}(E) \de E 
 &= k_{\rm B}^2 (T_L-T_R) \,x_0
 \left[ - \ln(2) \; T_R +T_L (\ln(1+\e^{-x_1}) - \ln(1+\e^{-x_0}))
 \right] \\
 &\phantom{=} + k_{\rm B}^2(T_L^2 - T_R^2) \Li_2(-\e^{-x_0}) -
 k_{\rm B}^2T_L^2 \Li_2(-\e^{-x_1}) - \frac{\pi^2}{12} k_{\rm B}^2T_R^2 \ ,
 \end{split}
\end{equation}
as reported in the main text.
The same procedure gives for the right current
\begin{equation}
 \label{sm:eq:IQR:refr}
 \begin{split}
 \int_{\mu_r}^{\epsilon_0} (E-\mu_R) f_{LR}(E) \de E & =
 (\epsilon_0-\mu_R)(g_L(\epsilon_0) - g_R(\epsilon_0))
 \\
 &\phantom{=} + \beta_L^{-2} [ \Li_2(-\e^{-x_0}) - \Li_2(-\e^{-x_1})]
 -\beta_R^{-2} [\Li_2(-\e^{-x_0}) - \Li_2(-1)] \\
 &= - k_{\rm B}^2 T_R (T_L-T_R) \,x_0 \ln(1+\e^{-x_0}) \\
 &\phantom{=} + k_{\rm B}^2(T_L^2 - T_R^2) \Li_2(-\e^{-x_0}) -
 k_{\rm B}^2T_L^2 \Li_2(-\e^{-x_1}) - \frac{\pi^2}{12} k_{\rm B}^2T_R^2 \ .
\end{split}
\end{equation}
For the input power, the integral is 
\begin{equation}
 \label{sm:eq:Pin:refr}
 \begin{split}
 (\mu_R - \mu_L) \int_{\mu_r}^{\epsilon_0} f_{LR}(E) \de E
 & = (\mu_R - \mu_L)( g_{LR}(\epsilon_0) - g_{LR}(\mu_R) ) \\
 & = k_{\rm B}^2 (T_L - T_R) x_0
 \left[ - \ln(2)\, T_R + (T_L - T_R) \ln(1+\e^{-x_0}) + T_L \ln(1+\e^{-x_1})
 \right] \ .
\end{split}
\end{equation}

\section{The low power regime}
\label{AppC:infinitesimal:integral}
In this regime the integration limits are very close.  This means that
we  can  use  the  Taylor  expansion of  the  $g$  and  $h$  functions
introduced          in         \eqref{sm:eq:primitive:ff}          and
\eqref{sm:eq:primitive:Eff}.  The  procedure is quite similar  for all
the needed  integrals, so we present  some details only for  one case.
For instance, for the power in the thermal machine mode we need
\begin{equation}
 \label{sm:eq:P:tm:low}
 \begin{split}
 \int_{\epsilon_0}^{\epsilon_0 + \Delta} f_{LR}(E) \de E
 & = g_{LR}(\epsilon_0 + \Delta) - g_{LR}(\Delta) \\
 &= g_{LR}^{\prime}(\epsilon_0) \Delta +
 g_{LR}^{''}(\epsilon_0) \frac{\Delta^2}{2} +
 g_{LR}^{'''}(\epsilon_0) \frac{\Delta^3}{6} + O(\Delta^4)\\
 & = f_{LR}(\epsilon_0) \Delta +
 f_{LR}^{'}(\epsilon_0) \frac{\Delta^2}{2} +
 f_{LR}^{''}(\epsilon_0) \frac{\Delta^3}{6} + O(\Delta^4) \ .
\end{split}
\end{equation}
Now using the fact that $f_{LR}(\epsilon_0) = 0$ and the function $\psi(x)=-1/(\e^x+1)$ as defined in the main text, we find
\begin{align}
 \label{sm:eq:fLR:d1d2}
 f_{LR}^{'}(\epsilon_0) &= (\beta_L - \beta_R) \frac{-\e^{x_0}}{(\e^{x_0}+1)^2} = \frac{T_L- T_R}{k_{\rm B}T_L T_R} \, \psi^{'}(x_0) \\
 f_{LR}^{''}(\epsilon_0) &= (\beta_L^2 - \beta_R^2) \frac{\e^{x_0}(\e^{x_0}-1)}{(\e^{x_0}+1)^3} = \frac{T_L^2- T_R^2}{k_{\rm B}^2T_L^2 T_R^2} \, \psi^{''}(x_0) 
\end{align}
and finally
\begin{equation}
 \label{sm:eq:P:tm:low:fin}
 \begin{split}
 \int_{\epsilon_0}^{\epsilon_0 + \Delta} f_{LR}(E) \de E
 & = \frac{ T_L- T_R}{2 k_{\rm B}T_L T_R} \, \psi^{'}(x_0) \Delta^2 +
 \frac{T_L^2- T_R^2}{6 k_{\rm B}^2T_L^2 T_R^2} \, \psi^{''}(x_0) \Delta^3 + O(\Delta^4)
 \end{split}
\end{equation}
as used in \eqref{eq:Poutl}.

\bibliography{bibliography} 

\end{document}